\documentclass[%
 reprint,
 amsmath,amssymb,
 aps,
pra,
]{revtex4-2}
\usepackage{xcolor}
\usepackage{graphicx}% Include figure files
\usepackage{physics}
\usepackage{floatrow}
\usepackage{subfig}
\usepackage{dcolumn}% Align table columns on decimal point
\usepackage{bm}% bold math
\usepackage{cleveref}
\usepackage{caption}
\begin{document}

%\preprint{APS/123-QED}

\title{Investigating ferromagnetic response in monolayer CVD grown  MoS$_{2}$ flakes using quantum weak measurement}% Force line breaks with \\

\author{Wardah Mahmood}\thanks{Equally contributing}
\affiliation{Department of Physics, Syed Babar Ali School of Science and Engineering, Lahore University of Management Sciences (LUMS), Opposite Sector U, DHA, Lahore 54792, Pakistan.}
\author{Muhammad Hammad Raza Gardezi}\thanks{Equally contributing}
\affiliation{Department of Physics, Syed Babar Ali School of Science and Engineering, Lahore University of Management Sciences (LUMS), Opposite Sector U, DHA, Lahore 54792, Pakistan.}
\author{Muhammad Arshad} \affiliation{Nanoscience and Technology Department, National Centre for Physics, Quaid-i-Azam University Campus, Islamabad.}
\author{Muddasir Naeem}
\affiliation{Department of Physics, Syed Babar Ali School of Science and Engineering, Lahore University of Management Sciences (LUMS), Opposite Sector U, DHA, Lahore 54792, Pakistan.}
\author{Ammar Ahmed Khan}
\affiliation{Department of Physics, Syed Babar Ali School of Science and Engineering, Lahore University of Management Sciences (LUMS), Opposite Sector U, DHA, Lahore 54792, Pakistan.}
\author{Syed Adnan Raza}
\author{Muhammad Sabieh Anwar}%
 \email{sabieh@lums.edu.pk}
\affiliation{Department of Physics, Syed Babar Ali School of Science and Engineering, Lahore University of Management Sciences (LUMS), Opposite Sector U, DHA, Lahore 54792, Pakistan.
\\}%

\date{\today}% It is always \today, today,

\begin{abstract}
We synthesize  MoS$_{2}$ atomic layer flakes at different growth conditions to tailor S-terminated and Mo-terminated edge defect states that are investigated for their ferromagnetic response. We leverage quantum weak measurement principles to construct a spin Hall effect of light-based magneto-optic Kerr effect (SHEL-MOKE) setup to sense the ultra-small magnetic response from the synthesized atomic layers. Our findings demonstrate that Mo-terminated edge states are the primary source of ferromagnetic response from  MoS$_{2}$ flakes, which is consistent with X-ray photoelectron, Raman and photoluminescence spectroscopic results. In the process, we demonstrate SHEL-MOKE to be a robust technique to investigate ultra weak properties in novel atomic-scale materials.
\end{abstract}

\maketitle

\section{\label{sec:level1}Introduction}

Two-dimensional (2D) materials have become central to the field of spintronics. Numerous studies reveal that semiconducting 2D materials display superior spin transport properties compared to their metallic counterparts \cite{li2014electrical,tian2014topological}.  Transition metal dichalcogenides (TMDs) are an interesting class of materials that show unique thickness dependent electronic and optical properties. Among these compounds, molybdenum disulfide (MoS$_{2}$) is a highly stable material possessing a number of interesting properties, including relatively high carrier mobility \cite{mir2020recent}, a moderate layer-dependent electronic band gap \cite{splendiani2010emerging}, and strong spin-orbit coupling \cite{fan2016valence}. Prior attempts at  extrinsic ferromagnetism primarily involved either creating spin polarization through doping or defect engineering, mostly in bulk MoS$_{2}$.  In this direction a significant amount of work has been done to produce dilute magnetic semiconductors (DMS) by substituting Mo with transition metal elements. Sanikop \emph{et al.} compared the ferromagnetic response in Co and Mn doped bulk MoS$_{2}$, concluding that both elements enhanced ferromagnetism when doped in small quantities. However, at higher doping levels, Mn substitution of Mo became inefficient, while Co doping lead to the segregation of an antiferromagnetic phase, resulting in paramagnetic behavior \cite{SANIKOP2021168226}. Fu \textit{et al}. synthesized Fe doped monolayer MoS$_{2}$ that demonstrated ferromagnetic behavior \cite{fu2020enabling}. However, precise control over doping and the resulting crystal structure remained a significant challenge. In addition to imparting a magnetic response, dopants are also expected to affect the electronic structure, which may be undesirable for a given application.

Recently, ferromagnetism in pure MoS$_{2}$ single crystals was investigated after annealing the material at high temperatures, resulting in an increase in the amount of S vacancies \cite{park2024strong}. Similarly, another study reported ferromagnetism in MoS$_{2}$ nanosheets by generation of S vacancies through a hydrothermal synthesis method, which resulted in localized transformation of 1H MoS$_{2}$ to 1T MoS$_{2}$ within the lattice  \cite{cai2015vacancy}. In both works, the increased ferromagnetism was linked to exchange of these interaction between the S vacancies and the Mo\textsuperscript{4+} state. However, these investigations were limited to bulk MoS$_{2}$ and did not explore atomic layers which are more viable and important for spintronics applications due to their direct band gap \cite{mak2010atomically,wang2017defects}.

Experimental studies on spontaneous magnetism in pristine MoS$_{2}$ are limited. Tongay \emph{et al.} obtained ferromagnetic signal from bulk MoS$_{2}$ single crystals and attributed the phenomenon to zigzag edges at the grain boundaries \cite{tongay2012magnetic}. Similarly, Zhou \emph{et al.} studied ferromagnetism in CVD-grown MoS$_{2}$ flakes and pyramids via vibrating sample magnetometry (VSM), and their study of pyramid structures using magnetic force microscopy (MFM) showed phase contrast indicating the origin of ferromagnetism in MoS$_{2}$ flakes to indeed be from the edge states. However, while these authors detected ferromagnetism in MoS$_{2}$ pyramids, they were unable to discover this behavior in monolayer flakes \cite{zhou2018robust}. Thus, it becomes evident that more sensitive experimental techniques may be required to overcome the limitations of traditional instruments and conventional techniques to measure the magnetic behavior in such materials at single-layer atomic thickness.

Side by side with experimental investigation, theoretical studies on ferromagnetism in MoS$_{2}$ have indicated the existence of intrinsic magnetism resulting from edge states \cite{zhou2018robust}. Depending on the type of edge, the atoms present in the bulk material have different coordination compared to the edge atoms. Mo-zigzag terminated edges in atomic layers consist of Mo with 4-fold coordination, compared to 6-fold coordination present within the bulk of flake, normally showing a diamagnetic behavior. The 4d orbitals of Mo within the bulk of the flake are spin-paired under 6-fold coordination due to ligand field splitting. In comparison, the unsaturated Mo at the edges of the flake show a spin polarized 4d orbital. Similarly, unsaturated S atoms in the S-terminating edges result in spin polarization in the 3p orbital \cite{zhang2007magnetic}. On the other hand, armchair edges produce no magnetic moment as the S and Mo atoms alternate along the edge resulting in absence of unpaired electrons. Therefore, in order to study magnetic behavior in MoS$_{2}$, it is imperative to synthesize flakes with high density of zigzag edges.

In this study, we examine ferromagnetism in monolayer CVD grown MoS$_{2}$ flakes using quantum weak value measurement. The weak value based quantum measurement technique, originally formulated within quantum mechanics, provides a tool to amplify and probe small physical changes that otherwise pose a challenge to detection \cite{xu2024progress}. Weak measurement theory proposes that a quantum observable can be weakly coupled to a measuring device, leading to an amplified shift in the measuring device’s pointer reading. Unlike a strong measurement, weak measurement does not collapse the wave function and allows for projection of state evolution on a post-selected state. This is where the most dramatic effects emerge with notably ultra large deviations in the pointer reading, far surpassing the range of allowed eigenvalues of the quantum state \cite{aharonov1988result}. The first experimental realization of weak measurement in the field of optics was demonstrated in 1989 by Duck \emph{et al.} providing experimental conditions required to access “weak values” \cite{duck1989sense}. Since then, weak value amplification has been exploited to observe beam splitting like the spin Hall effect of light (SHEL), a phenomenon where a linearly polarized state of light splits into right and left circularly polarized light after interacting with a birefringent surface. This separation in space is extremely sensitive to changes in polarization of the incident state \cite{qin2011observation}. Other examples of mechanical spatial beam shifts include Goos–Hänchen and Imbert–Fedorov \cite{goswami2014simultaneous}. These ultrasensitive sensing techniques have been pivotal in enriching our understanding of novel phenomena in nanoscale materials \cite{li2020measurement}.

Measuring ferromagnetic response, especially from atomic layer structures poses a significant challenge due to the weak signal strength. Conventional techniques like superconducting quantum interference device (SQUID magnetometry) and magneto-optical Kerr effect (MOKE) setups struggle to detect minute polarization changes as the signal drops out of range of typical detectors \cite{cai2022room}. In this work, we leverage SHEL combined with quantum weak value amplification, enabling detection of such magnetic responses that would otherwise remain inaccessible.

This article discusses details of the synthesis of monolayer MoS$_2$ flakes under varying growth conditions, highlighting the impact of growth temperature, gas flow rate and substrate position relative to precursors, on the defect state and edge state density. We present details on morphology and compositional variations across different samples. An introduction to quantum weak values for optical metrology includes the mathematical formalism, followed by an explanation of the experimental implementation of weak value measurement based SHEL-MOKE. Detailed spectroscopic analyses, including Raman, photoluminescence (PL), and X-ray photoelectron spectroscopy (XPS) measurements are employed to characterize the defect states, revealing their role as polarization-altering active sites. These states are detected through their ferromagnetic hysteresis response enabled by SHEL-MOKE. Finally, we conclude with an analysis on how the density of polarization-altering defect states influences ferromagnetic characteristics.

\section{\label{sec:level1}Experimental methods}

\subsection{\label{sec:level2}Growth of monolayer MoS$_{2}$ flakes using chemical vapor deposition}
\begin{figure}
 \centering
 \includegraphics[width=1\linewidth]{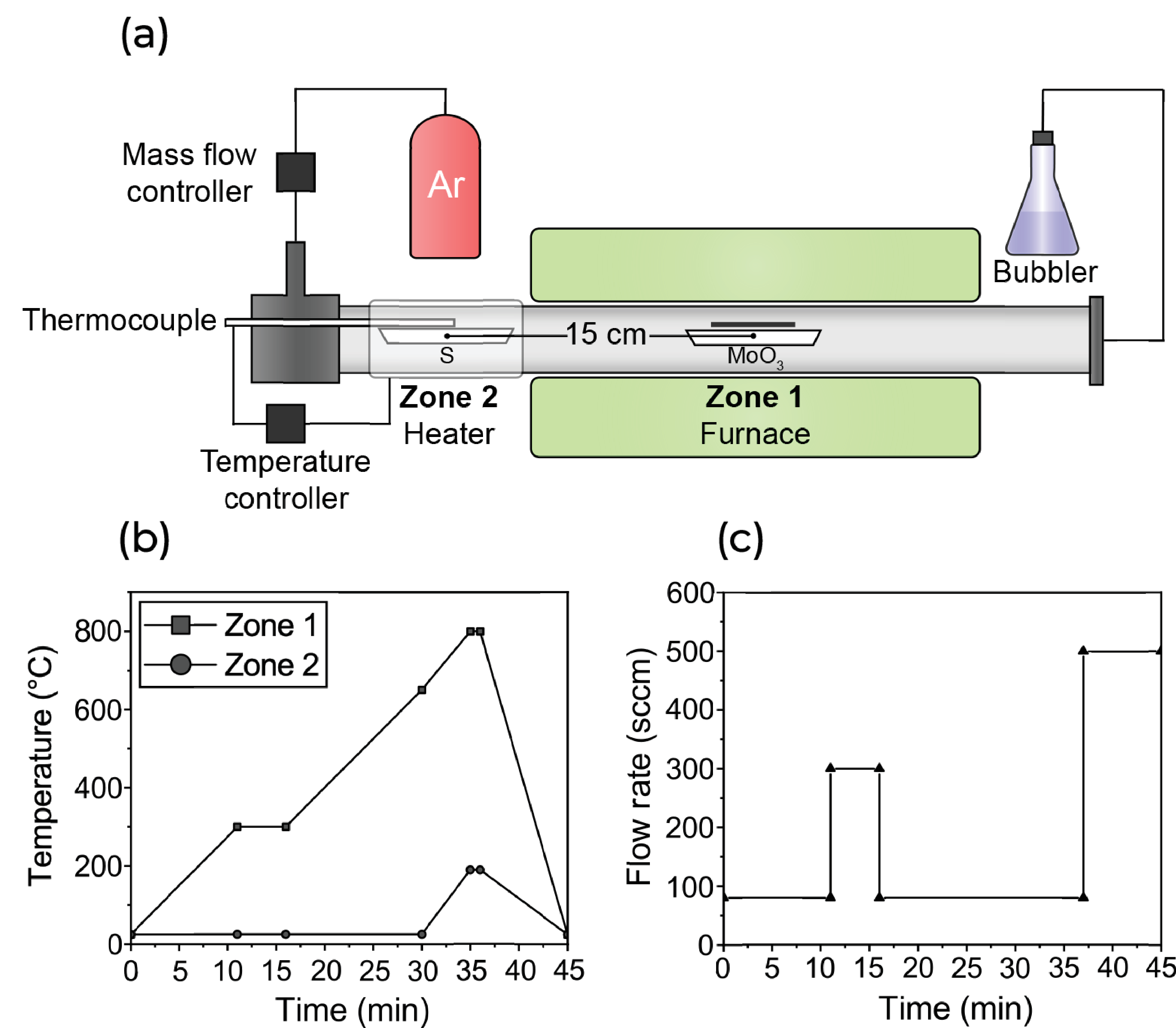}
 \caption{(a) Illustration of the CVD setup. (b) Temperature program of the heating zones. (c) Gas flow profile for the system.} \label{schematic CVD}
\end{figure}
The setup  and profiling used for the growth of monolayer MoS$_{2}$ is shown in Figure$~$\ref{schematic CVD}. A dual zone homemade tube furnace is used with a quartz tube of inner diameter $1~$inch. Within the tube, 1 mg of molybdenum trioxide (MoO$_{3}$) powder was placed downstream in a quartz boat at the center of the furnace in Zone 1, while a second quartz boat containing 10 mg of sulfur (S) was placed in Zone 2 which is controlled via a separate set of heating coils. A silicon dioxide (SiO$_{2}$, $300~$nm) substrate was placed facing down above the MoO$_{3}$ powder. Before loading, the substrate was cleaned in an ultrasonic bath using acetone, isopropanol alcohol (IPA) and methanol, while rinsing with DI water between each cycle. Zone 1 was heated up from room temperature to $300^{\circ}$C at a rate of $25^{\circ}$C/min under a flow of Ar gas at $40~$sccm, $80~$sccm and $120~$sccm. The furnace was held at $300^{\circ}$C for 5 mins, during which the Ar gas flow was increased to $300~$sccm to purge out impurities. Subsequently, the furnace was made to reach $800^{\circ}$C in 19 mins (and 15 mins for growth at $700^{\circ}$C). The heater in Zone 2 was switched on when Zone 1 attained a temperature of $650^{\circ}$C (and $550^{\circ}$C for growth at $700^{\circ}$C). Zone 2 was allowed to reach $190^{\circ}$C in approximately 6 mins, which coincided with Zone 1 reaching the set temperature. When Zone 1 achieved the set temperature of $800^{\circ}$C (or $700^{\circ}$C), the allowed growth time was 1 min. After the growth time finished, system was purged with Ar and the reaction was quenched by lifting up the furnace hood of Zone 1, and simultaneously switching off the heater of Zone 2. The optimized parameters for temperature, gas flow rates, and substrate placement ensure controlled growth of high-quality monolayer MoS$_{2}$ flakes suitable for detailed morphological and magnetic studies.

\subsection{\label{sec:level2}Quantum weak value measurement}
Weak value amplification leverages the nearly orthogonal pre- and post-selected states to enhance the optical centroid shift of a Gaussian optical beam reflecting from a magnetic surface. The shift is directly related to spin-dependent interactions within a magnetic material. We measure this centroid shift in order to ascertain the ferromagnetic response of atomic layer MoS$_{2}$. The shift is sensitive to the spin orbit coupling and anisotropies within the material, and a hysteresis-like profile of the centroid shift is interpreted as a signature for edge-induced ferromagnetism.

\begin{figure}[h]
    \centering
    \includegraphics[width=1\linewidth]{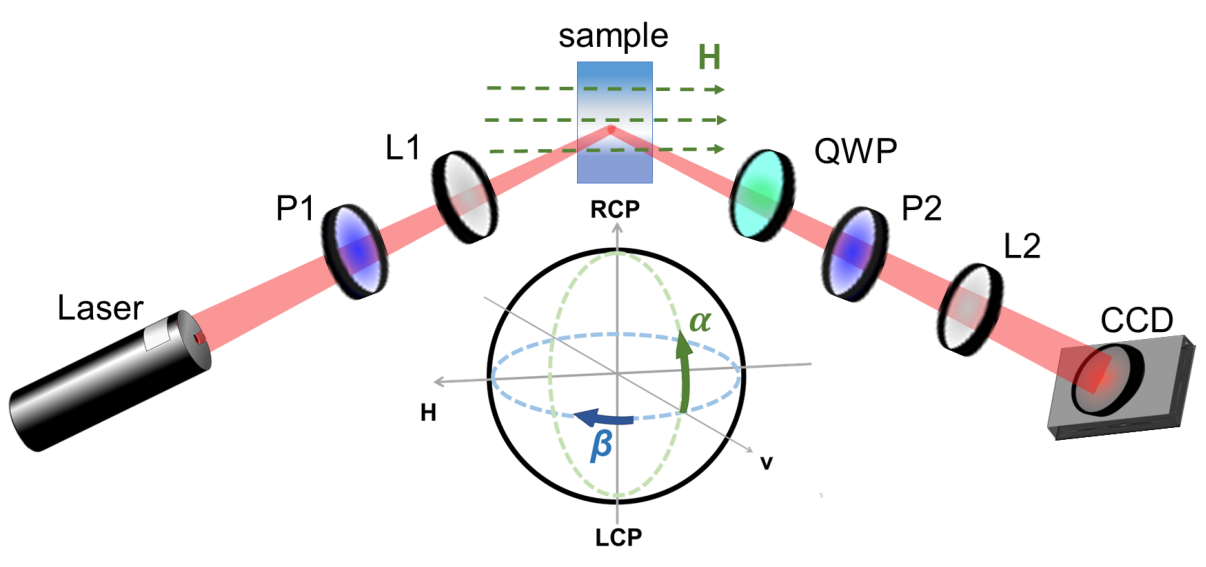}
    \caption{Setup illustration to measure SHEL-MOKE from a sample placed in an external magnetic field and probed in longitudinal configuration. The lens (L1) focuses beam onto the sample, polarizer (P1) is used to prepare $|\psi_{pre}\rangle$, analyzer (P2) and a quarter wave plate (QWP) are used to adjust angles $\beta$ and $\alpha$ respectively, thus preparing $|\psi_{post}\rangle$. Using a second lens (L2), the final state is focused onto a camera for imaging. The Poincar$\Acute{\text{e}}$ sphere shows the action of QWP and P2 on creating polarization state  $|\psi_{post}\rangle$. The angle $\beta$ moves the state along the H-V polarization axis, and $\alpha$ moves the state along circular polarization axis.}
    \label{schematic SHEL-MOKE}
\end{figure}

A HeNe $632~$nm continuous wave laser is prepared in an initial state $|\psi_{initial}\rangle$ that jointly describes the polarization and spatial beam profile of the laser which has a Gaussian transverse profile

\begin{equation}
|\psi_{initial}\rangle=|\psi_{pre}\rangle \otimes \phi_{k},
\end{equation}

\begin{align*}
&|\psi_{pre}\rangle = \begin{bmatrix}
      \cos{\theta_o} \\
      \sin{\theta_o}
\end{bmatrix}, \\
&\phi_k = \frac{w_o}{\sqrt{2\pi}} \exp\left(-\left(\frac{i z}{2k_o}
+ \frac{w_o^2}{4}\right)(k_x^2 + k_y^2)\right).
\end{align*}

\begin{figure}[h]
    \centering
    \includegraphics[width=1\linewidth]{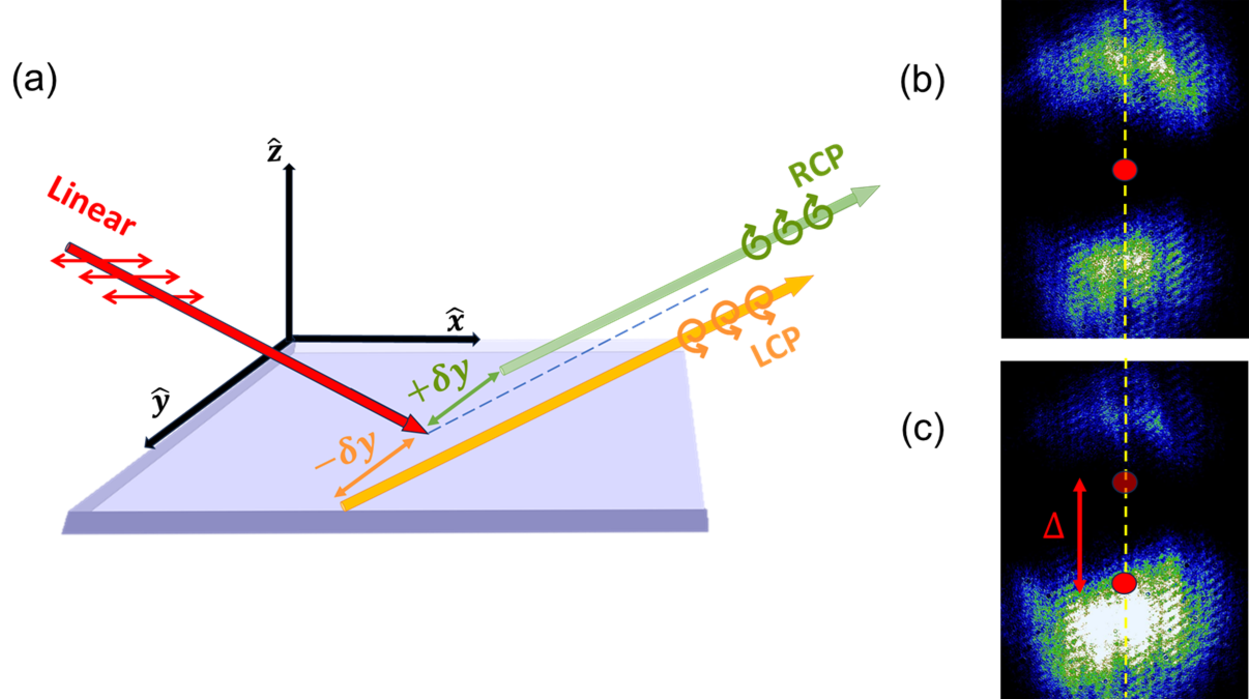}
    \caption{(a) The spin Hall effect of light (SHEL) phenomenon. A linearly polarized beam of light passes through an optical interface. The resulting interaction splits the incident light into its circular components, RCP and LCP, displaced by $- \delta y$ and $+\delta y$ respectively. (b) Experimental observation of this splitting in the absence of any external applied field is captured on a CCD. The centroid position is represented by a red dot along $\hat{y}$. (c) Significant centroid shift ($\Delta$) is observed in presence of external applied magnetic field on the sample.}
    \label{schematic SHEL-MOKE and centroid}
\end{figure}

\noindent In the above expressions $w_o$ is the beam waist, $z$ is the propagation length,  and $k_o=2\pi/\lambda$ is the wave vector along $\hat{z}$. The initial polarization is aligned at an angle $\theta_o$ with respect to the horizontal axis in the lab frame. The pre-selected polarization state $|\psi_{pre}\rangle$ is prepared using polarizer (P1) and the beam profile is shaped using a spatial filtering system. The scheme of this experiment is illustrated in Figure~\ref{schematic SHEL-MOKE}. We choose horizontal polarization as our pre-selected state which can be conveniently decomposed in the circular basis as $|\psi_{pre}\rangle= |H\rangle =(|+\rangle+|-\rangle)/\sqrt{2}$. This is focused on the sample using lens (L1) at the Brewster angle in order to maximize the amplitude of beneficial Fresnel coefficients. After reflection from the sample, the state evolves and undergoes the SHEL phenomenon, splitting into right circularly polarized (RCP) and left circularly polarized (LCP) components along the $\hat{y}$ direction. The centroid position is the spatial position in the xy plane that corresponds to the weighted average intensity from the RCP and LCP. As the interaction changes due to an external applied magnetic field, the weighted average intensity shifts along the $\hat{y}$ from it's initial set point. This is defined as the centroid shift $\Delta$. The geometrical setting can be seen in Figure~\ref{schematic SHEL-MOKE and centroid}.

Following the comprehensive theoretical insights from the work of Panda \emph{et al.}, we evaluate the optical response of this measurement setup \cite{panda2022high}. The evolved state after the optical interaction can be expressed as $|\psi_{evl}\rangle_{H}=R_{c}|H\rangle$, where $R_{c}$ is the general reflection matrix written in circular basis \cite{luo2011enhancing},
\begin{equation}
 R_{c}=\begin{pmatrix} R_{c}^{11} & R_{c}^{12} \\ R_{c}^{21} & R_{c}^{22}
 \label{Rc matrix}
 \end{pmatrix},
\end{equation}

\begin{align*}
&R_{c}^{11}=r_{o}^{pp} \left[1 + i k_{y}^{r} \delta_H - k_{x}^{r} \zeta_H - i \theta_H (1 - k_{x}^{r} \zeta_H^{\prime} + i k_{y}^{r} \delta_H')\right]\\
&R_{c}^{12}=-ir_{o}^{ss} \left[1 + i  k_{y}^{r} \delta_V - k_{x}^{r} \zeta_V + i \theta_V (1 - k_{x}^{r} \zeta_V' + i k_{y}^{r} \delta_V')\right]\\
&R_{c}^{21}= r_{o}^{pp} \left[1 - i k_{y}^{r} \delta_H - k_{x}^{r} \zeta_H + i \theta_H (1 - k_{x}^{r} \zeta_H - i k_{y}^{r} \delta_H')\right]\\
&R_{c}^{22}=ir_{o}^{ss} \left[1 - i k_{y}^{r} \delta_V - k_{y}^{r} \zeta_V - i \theta_V (1 - k_{x}^{r} \zeta_V' - i k_{y}^{r} \delta_V')\right].
\end{align*}

\noindent The reflection matrix $R_{c}$ not only takes care of reflection components, but also transforms the linear input basis to the circular basis. Here, we have made the substitutions

\begin{eqnarray}
\delta_H &=& \frac{\cot \theta_i}{k_o} \left(1 + \frac{r_o^{ss}}{r_o^{pp}}\right),
\delta'_H =  \frac{\cot \theta_i}{k_o} \left(1 - \frac{r_o^{ps}}{r_o^{sp}}\right), \\
\delta_V &=&  \frac{\cot \theta_i}{k_o}\left(1 + \frac{r_o^{pp}}{r_o^{ss}}\right),
\delta'_V =  \frac{\cot \theta_i}{k_o} \left(1 - \frac{r_o^{sp}}{r_o^{ps}}\right), \\
\zeta_H &=& \frac{1}{k_o r_o^{pp}} \frac{\partial r_o^{pp}}{\partial \theta_i},
\zeta'_H = \frac{1}{k_o r_o^{sp}} \frac{\partial r_o^{sp}}{\partial \theta_i}, \\
\zeta_V &=& \frac{1}{k_o r_o^{ss}} \frac{\partial r_o^{ss}}{\partial \theta_i},
\zeta_V' = \frac{1}{k_o r_o^{ps}} \frac{\partial r_o^{ps}}{\partial \theta_i},\\
\theta_H &=& \frac{r_o^{sp}}{r_o^{pp}}, \theta_V=\frac{r_o^{ps}}{r_o^{ss}}.
\end{eqnarray}

The Fresnel coefficients $r_o^{ss}, r_o^{pp},r_o^{sp}$ and $r_o^{ps}$ are Taylor expanded to first order approximation considering the spread of the plane wave vectors in the paraxial regime. In Equation \eqref{Rc matrix} the term $(1 \pm i k_y^r\delta_{(H,V)})$ represents the spin-orbit interaction (SOI) due to the diagonal Fresnel reflection coefficients. Further, the terms $\theta_{H,V}$ represent the optical activity of the material and can be read as perturbation to the original SHEL shift. For magnetic materials, the term $\theta_{H}$ and  $\theta_{V}$ represents the Kerr effect for $H$ and $V$ polarized incident light.
Finally the beam passes through a combination of a quarter wave plate (QWP) and polarizer (P2) that allows post-selection onto the state represented by \cite{panda2022high},
\begin{equation}
|\psi_{post}\rangle_{H}= i\left[e^{i\beta}\cos({\frac{\pi}{4}+\alpha})|-\rangle-e^{-i\beta}\sin({\frac{\pi}{4}+\alpha})|+\rangle\right].
\label{postselection}
\end{equation}
In the preceding formula, all optical components are aligned with respect to the optical axis of polarizer (P1), which is locked in the horizontal polarization. In our definition $\alpha$ is the small angle deviation of QWP, and $\beta$ is the small angle deviation from orthogonal state of P1. Hence the final state after reflection can be observed as a projection on the post-selected state, i.e., $|\psi_{\text{final}} \rangle_{H}=|\psi_{post}\rangle_{H}\langle\psi_{post}|\psi_{evl}\rangle_{H}$, which finally yields
\begin{equation}
\begin{aligned}
\lvert \psi_{final}& \rangle_{H}\\
 = ~ & \gamma_{H} \left(1 + i k_y^r \delta_H A_{H}^W - k_x^r \zeta_{H} \right) -\\
& i \theta_{H} \left(A_{H}^W - k_x^r \zeta_H' A_{H}^W - i k_y^r \delta_H' \right) \lvert \psi_{post} \rangle_{H} \langle \psi_{post} \lvert \psi_{pre} \rangle_{H},
\end{aligned}
\end{equation}

\noindent where we have the operator for SHEL $\hat{A} = \begin{pmatrix} 1 & 0 \\ 0 & -1 \end{pmatrix}$ and the so called weak value $A_H^W$ is expressed as

\begin{equation}
    A^{W}_{H} = \frac{_H\langle \psi_{post} | \hat{A} | \psi_{pre} \rangle_{H}}{_H\langle \psi_{post} | \psi_{pre} \rangle_{H}} \approx \frac{ \sin 2\alpha + i \sin 2\beta \cos 2\alpha}{2 \left| _H\langle \psi_{post} | \psi_{pre} \rangle_{H} \right|^2}.
\end{equation}

\noindent In order to observe this in position coordinates, we Fourier transform the final wave function bringing this to positional variables
\begin{equation}
    |\Psi_{final} \rangle_{H} = \iint_{-\infty}^{\infty} \exp \left\{ i \left( k_x x + k_y y + k_o z \right) \right\} |\psi_{final} \rangle_{H} \phi_k \, dk_x \, dk_y.
\end{equation}
 The shift in centroid position ($\Delta$) of the splitting along $\hat{y}$ is calculated using
\begin{align}
\Delta y^{H} = \frac{\langle \Psi_{final} | \hat{y} | \Psi_{final} \rangle_{H} } {\langle \Psi_{final} | \Psi_{final} \rangle_{H}} =\frac{\Delta y^{N}}{\Delta y^{D}},
\end{align}
where the numerator $\Delta y^N$ and denominator $\Delta y^D$ are spelled out as,

\begin{widetext}
\begin{eqnarray}
\Delta y^N &=& (a^2 + b^2) [\frac{1}{2a - 2ib}(i \delta_H A_w - \theta_H \delta_H^\prime-\theta_H^* \delta_H A_w A_w^* - i \theta_H \theta_H^* \delta_H^\prime A_w^*) \nonumber\\
& +&   \frac{1}{2a + 2ib}(-i \delta_H^* A_w^* - \theta_H  \delta_H^* A_w A_w^* - \theta_H^* \delta_H^{\prime*} + i \theta_H \theta_H^* \delta_H^{\prime*} A_w)]\\
\Delta y^D &=& 1+i \theta_H^* A_w^*- i \theta_H A_w + \theta_H \theta_H^* A_w A_w^* \nonumber\\
& +& \frac{1}{4b} ( \zeta_H \zeta_H^*  + i \zeta_H \theta_H^* \zeta_H^{\prime*} A_w^* - i \theta_H \zeta_H^* \zeta_H^\prime A_w + \theta_H\theta_H^* \zeta_H^{\prime} \zeta_H^{\prime*} A_w A_w^* \nonumber \\
&+& \delta_H \delta_H^* A_w A_w^*  - i \delta_H \theta_H^* \delta_H^{\prime*} A_w + i \theta_H\delta_H^\prime\delta_H^* A_w^*+ \theta_H \theta_H^* \delta_H^\prime \delta_H^{\prime*}).
\end{eqnarray}
\end{widetext}

For simplicity $\Delta y^H$ is explained in terms of its expression in the numerator $\Delta y^N$ and the denominator $\Delta y^D$ where they are composed to their simplest form with constants $a=z/2k_o$ and $b=w_o^2/4$. These parameters highlight the influence of beam geometry and propagation length on the sensitivity of the setup measurement. The terms $\theta_H$, $\theta_H^*$  and their combinations represent the magneto-optical Kerr effect. The terms with $\theta_H^*\delta_H$ found in the numerator show that the polarization splitting is modulated by the material's magneto-optic properties. $A_w$ is the weak value of spin Hall operator which is highly sensitive to changes in polarization states mainly because they manipulate the angles $\alpha$ and $\beta$. The presence of this term scales the amplification of the centroid shift ($\Delta$), which is the key element for detecting ultra-small polairzation-dependent changes in our system. The terms including $\zeta_H\zeta_H^*$ and $\delta_H\delta_H^*$ quantify how spin-orbit coupling and spin-polarized effects in the material influence polarization based SHEL splitting.

This confirms that $\Delta y^H$ is an effective measure of magneto-optic phenomena. This well supports our understanding that the change in Kerr effect as a function of applied magnetic field would perturb the SHEL centroid $\Delta$ to shift if the material is magnetic in nature. Consequently, the scan of $\Delta$ versus magnetic field will elicit information about the ferromagnetism induced by the edges of the atomic MoS$_{2}$ crystals. Utilizing the sensitivity of SHEL-MOKE to magnetic changes in material, this behavior is studied and analyzed by adjusting appropriate pre-selected and post-selected states for each sample. Notably, the weak value technique detects a ferromagnetic signature with ultra-high sensitivity when a conventional technique such as the traditional MOKE is unable to detect a magnetic response.

\section{\label{sec:level1}Results and discussion}
\subsection{\label{sec:level2}Morphology control}
\begin{figure}[ht!]
    \centering
    \includegraphics[width=1\linewidth]{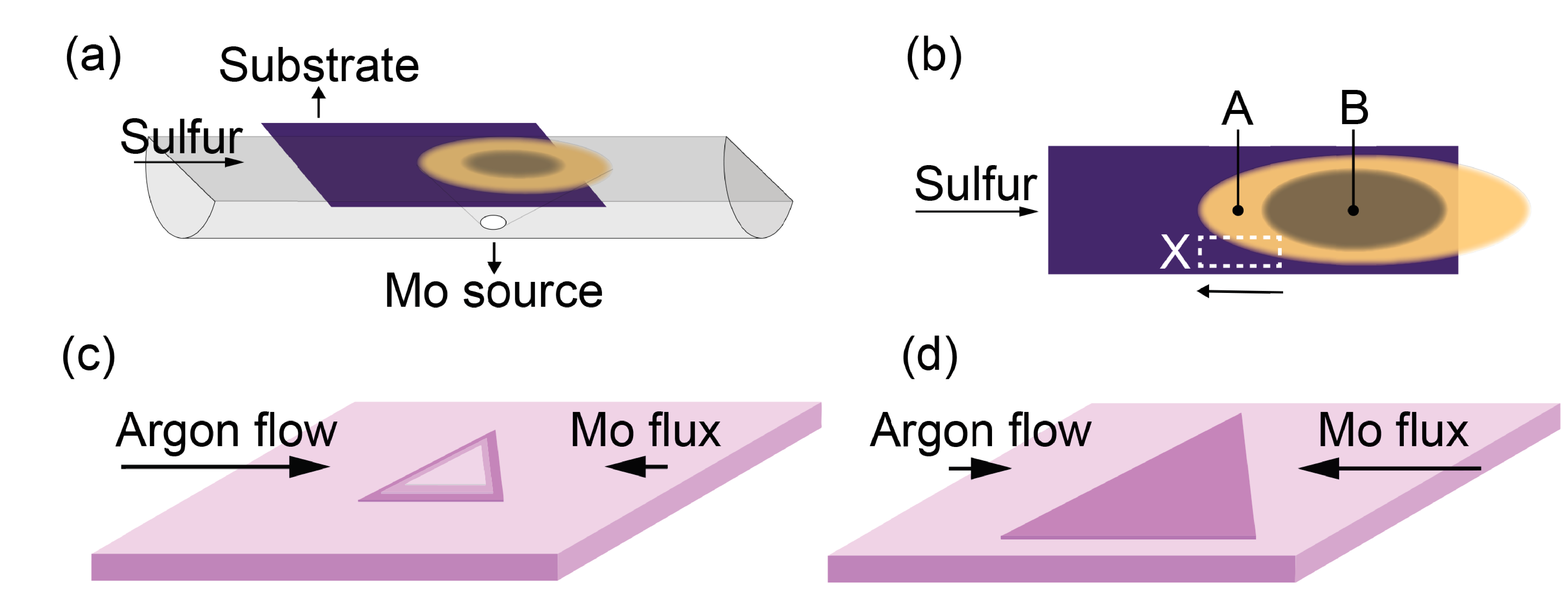}
    \caption{(a) Illustration of the substrate position SiO$_2$ (placed upside down) on quartz boat containing MoO$_{3}$ powder. (b) Representation of the deposition pattern observed on the front surface of substrate after growth. S:Mo vapor concentration increases in the direction of the black arrow across region 'X'. Domain growth of MoS$_2$ is illustrated at (c) high and (d) low argon gas flow rates. The Mo flux flows upstream against the argon flow.
}
    \label{schematic of substrate position and deposition pattern}
\end{figure}

The position of the substrate on the quartz boat relative to MoO$_{3}$ powder is illustrated in Figure~\ref{schematic of substrate position and deposition pattern} (a). For all experiments, the powder was placed at the center of the quartz boat, whereas the substrate was positioned slightly upstream to facilitate the study of deposition in that area. Typical growth zones are labeled in the schematic which correspond to the local S:Mo ratio.  Position ‘A’ indicates the region where monolayer MoS$_{2}$ is predominant, and position ‘B’ represents the area lying directly above the Mo source where bulk MoS$_{2}$ is found. The area and shape of the ellipse depend on the temperature around the Mo source and the gas flow rate inside the system. Through experimental trials, it was observed that the ellipse tends to decrease in size as the temperature is reduced; conversely, upon increasing the flow rate, it appears to skew in the direction of the gas flow, causing the region with flake growth to become narrower.

\begin{figure*} [ht!]
    \centering
    \includegraphics[width=0.7\linewidth]{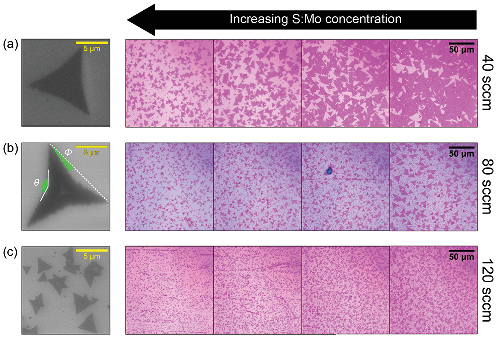}
    \caption{Optical microscope images showing the shape evolution of MoS$_{2}$ flakes across region ‘X’ grown at 700$^{\circ}$C at a Ar gas flow rate of (a) 40 sccm, (b) 80 sccm, and (c) 120 sccm. The scanning electron microscope (SEM) images reflect the shape evolution of flakes grown at (a) 40 sccm, (b) 80 sccm, and (c) 120 sccm. The SEM image on the left in (b) shows how edge angle ($\theta$) and vertex angle ($\phi$) are defined.}
    \label{optical microscope images of flakes 700}
\end{figure*}

\begin{figure*} [ht!]
    \centering
    \includegraphics[width=0.7\linewidth]{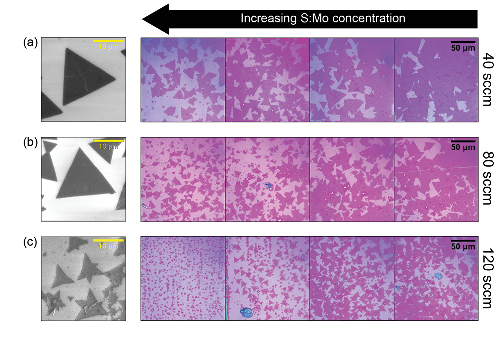}
    \caption{Optical microscope images showing the shape evolution of MoS$_{2}$ flakes across region ‘X’ grown at 800$^{\circ}$C at a Ar gas flow rate (a) 40 sccm, (b) 80 sccm, and (c) 120 sccm. The scanning electron microscope images reflect the shape evolution of flakes grown at (a) 40 sccm, (b) 80 sccm, and (c) 120 sccm.}
    \label{optical microscope images of flakes 800}
\end{figure*}

\begin{figure}[ht!]
    \centering
    \includegraphics[width=1\linewidth]{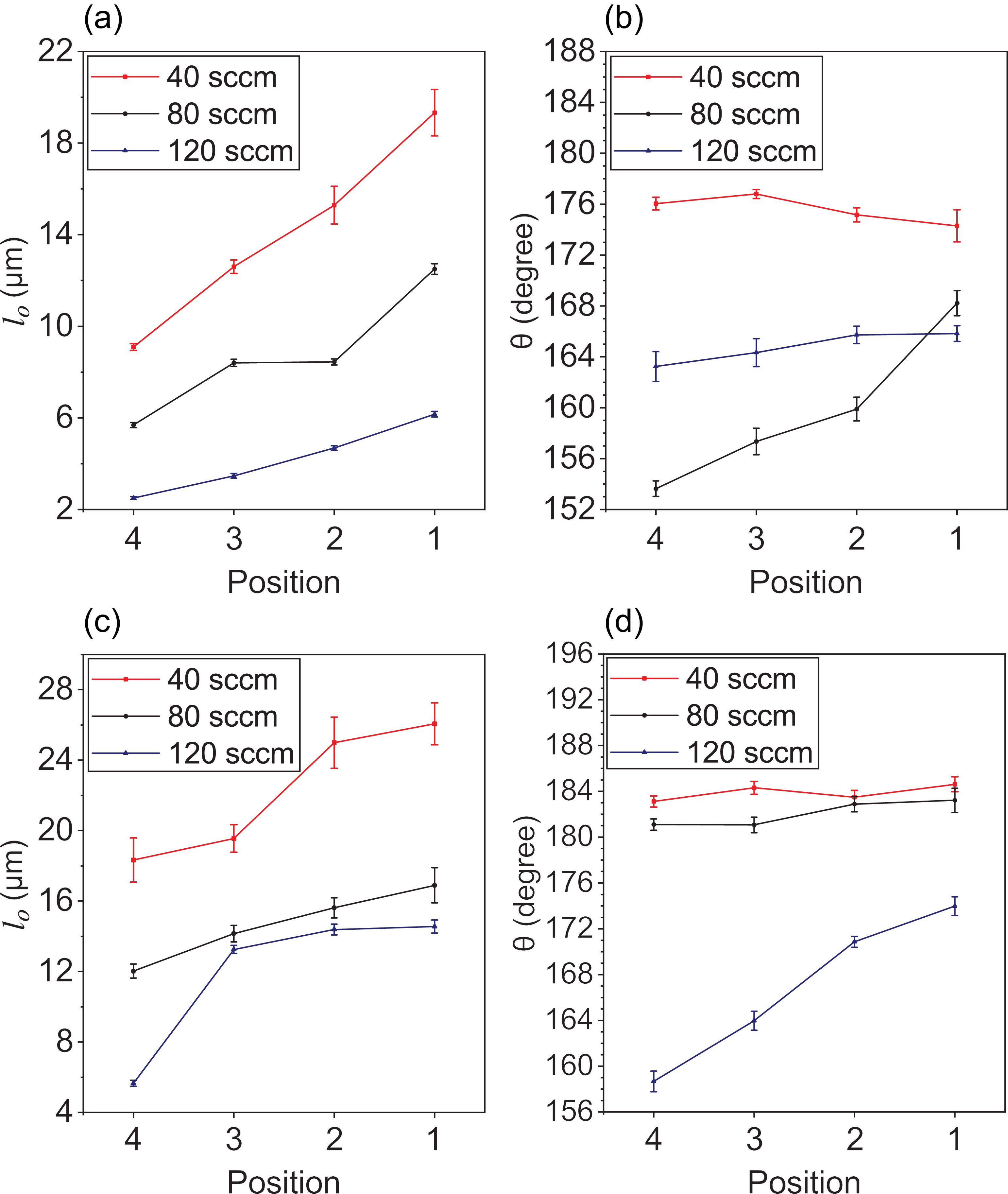}
    \caption{(a) Edge length ($l_o$) and (b) flake edge angle ($\theta$) for flakes across region ‘X’ grown at 700$^{\circ}$C with flow rate of $40~$sccm, $80~$sccm and $120~$sccm. (c) Edge length ($l_o$) and (d) edge angle ($\theta$) for flakes across region ‘X’ grown at 800$^{\circ}$C with flow rate of $40~$sccm, $80~$sccm and $120~$sccm.}
    \label{edge length and flake angle}
\end{figure}

The shape and size of the MoS$_{2}$ flakes were examined in the upstream region relative to the Mo source (denoted ‘X’ in Figure~\ref{schematic of substrate position and deposition pattern} (b) and is a subset of the region A, X$\subset$A) at Ar flow rates of $40~$sccm, $80~$sccm, and $120~$sccm. Figure~\ref{optical microscope images of flakes 700} and Figure~\ref{optical microscope images of flakes 800} show optical micrographs across region ‘X’, moving in the direction against the gas flow (right to left). Owing to the laminar flow in the CVD chamber at low flow rates, the distribution of S was relatively uniform away from the source. The Mo source, however, is positioned in close proximity to the substrate, thus the local concentration is primarily determined by the diffusion of the  MoO$_{3-x}$ vapor towards the substrate. The diffusion pattern is influenced by the temperature of the Mo precursor, the carrier gas flow rate and the distance from the source. This is reflected in the deposition pattern on the substrate which takes the shape of an ellipse, with the region directly above the source having the greatest amount of  MoO$_{3}$ concentration, and consequently the lowest S:Mo ratio. This results in the deposition of bulk MoS$_{2}$, unreacted rhombohedral MoO$_{2}$ and partially sulfurized MoO$_{x}$S$_{y}$ nanoparticles. To quantify the change in flake morphology across the region ‘X’, ImageJ software was used to measure and plot the mean edge length $(l_o)$ and edge angle $(\theta)$ of the MoS$_{2}$ flakes in each of the images. Figure~\ref{edge length and flake angle} (c-d) shows the variation in the edge length and edge angle of the flakes grown at 800$^{\circ}$C as a function of position when moving upstream from the Mo source.

At low gas flow rates, it was observed that the region in closest proximity to the Mo source exhibited triangular flakes with edges slightly bulging outwards. The lower gas flow rate allows for greater lateral diffusion of the MoO$_{3-x}$ reactant against the gas flow, creating a region of relatively low S:Mo concentration upstream. This phenomenon is reflected by the high mean edge angles within the range of 183$^{\circ}$ to 185$^{\circ}$, which remain approximately uniform across the area of interest with only a marginal change in the edge angle as the Mo concentration decreases. Lateral diffusion also facilitates the formation of seeding centers at a distance, thus promoting the growth of flakes at greater distances from each other.
It is well known that the edges of the MoS$_{2}$ flakes possess a higher surface free energy compared to the basal plane due to dangling bonds \cite{gaur2014surface}. The flakes preferentially grow in lateral direction along the higher energy plane provided that enough vapor is able to diffuse to the edges for growth. A low flow rate therefore enhances lateral growth at the flake edges while inhibiting bilayer nucleation on the flake surface facilitated by the increased lateral diffusion and mass transport upstream, refer to Figure~\ref{schematic of substrate position and deposition pattern} (c-d). Consequently, at $40~$sccm we observe large flakes with edge lengths exceeding 25$~\mu$m. Moreover, in the regions where the reactant supply is sufficient, the flakes grow laterally and become coalesced to form large area films. At distances further away from the Mo source along the gas flow direction, the flake size decreased to below 15$~\mu$m, despite the initial nucleation, because of the insufficient MoO$_{3-x}$.
Accordingly, it is noted that as the flow rate is increased to $80~$sccm, the flake sizes decrease. The mean edge length of the flakes closest to the Mo sources is 17$~\mu$m in the sample grown at $80~$sccm, which decreases to 12$~\mu$m in the fourth image (to the left). The edge angle is similar to the sample grown at $40~$sccm, which indicates that the S:Mo ratio remains low; however, it is evident from Figure~\ref{optical microscope images of flakes 800}, that the flake centers are relatively closer, suggesting that as the flow rate increases, the seeding centers nucleate in closer proximity to each other, resulting in the growth of smaller flakes.

At higher flow rates, the impact of gas flow on the flake shape is more pronounced. High flow rate inhibits the diffusion of MoO$_{3-x}$ against the direction of gas flow, creating a region of high S:Mo concentration in region ‘X’, see Figure~\ref{schematic of substrate position and deposition pattern} (b). Therefore, a sharp decline in the mean edge angle $(\theta)$ from 174$^{\circ}$ to 159$^{\circ}$ was observed at $120~$sccm as we moved against the flow of the gas. The Mo-zigzag edges reacted with the excess sulfur present, causing these edges to grow faster than the S-zigzag edges, thus leading to their disappearance over time. The amount of Mo able to diffuse against the flow in this instance decreases significantly relative to the previously discussed cases. This is further reflected by the sharp decline in the edge length as shown in Figure~\ref{edge length and flake angle}.

Likewise, we also compared the morphologies of flakes grown at 700$^{\circ}$C. Figure~\ref{edge length and flake angle} illustrates the variation in flake shape and size in relation to the changing S:Mo ratio across region ‘X’. The vapor pressure of MoO$_{3}$ decreases exponentially with temperature \cite{yang2018influences}. Consequently, at 700$^{\circ}$C, the amount of  vapor was expected to be significantly lower than that at 800$^{\circ}$C. As a result, the S:Mo ratio became even lower, resulting in curved-edge flakes growing at shorter distances from the Mo source. We observed a trend similar to that recorded at 800$^{\circ}$C, with the flakes grown at $40~$sccm exhibiting negligible change in the edge angle. However, the actual angles are much smaller in comparison, indicating the decreased partial pressure of the Mo precursor at low temperature. At $80~$sccm, the edge angles exhibited a more noticeable reduction, reaching a mean edge angle of 168$^{\circ}$ near the Mo source. As the S:Mo ratio increased when moving towards the S source, three-point star-shaped flakes with an average angle of 154$^{\circ}$ became apparent. In addition, a lower temperature also causes the MoO$_{3-x}$ vapor to diffuse over shorter distances, resulting in seeding centers forming in close proximity to each other. It can therefore be observed that the flakes are smaller in size than those grown at 800$^{\circ}$C, with significantly smaller flakes growing between the larger flakes.\\

At a flow rate of $120~$sccm, an anomalous behavior was observed, wherein the edge angle remained nearly constant across the substrate. It is noteworthy that as the flow rate increases, the region ‘X’ becomes narrower, bringing the study region closer to the Mo source which in turn results in a more uniform Mo concentration. Despite the presence of abundant S, the higher flow rate at low temperatures results in insufficient Mo for substantial growth at the Mo edges, thus inhibiting these edges from developing into sizes where large angles would manifest. In effect, the low lateral diffusion and high nucleation density limit the growth of these flakes.

\begin{figure}[ht!]
    \centering
    \includegraphics[width=1\linewidth]{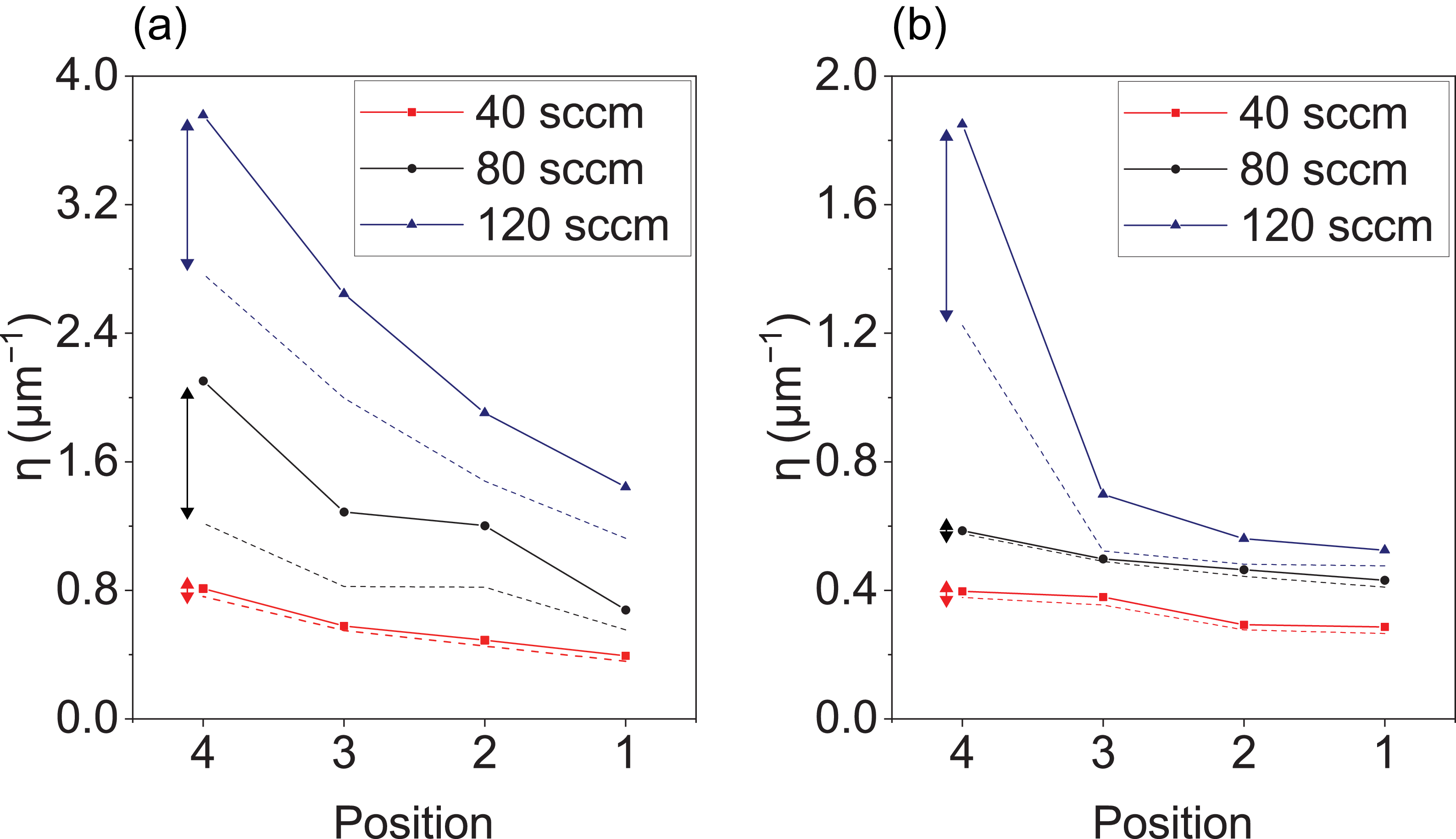}
    \caption{Perimeter-to-area ratio $(\eta)$ of MoS$_{2}$ flakes grown at (a) 700$^{\circ}$C and (b) 800$^{\circ}$C. The dotted lines represent the ratio calculated assuming flake edges to be straight for comparison. The double-sided arrow connects the respective $\eta$ to the corresponding straight flake edge.}
    \label{perimeter to area ratio 700 and 800}
\end{figure}

In order to achieve optimal conditions for invoking ferromagnetic response, it is essential to achieve maximum density of zigzag edges. One approach for this is to grow a high density of flakes close to each other, which inherently corresponds to a high initial nucleation density. As previously established, this would translate into a change in flake size, and hence, a change in both the area and edge length. Another factor that influences both the area and edge length is edge angle. A high edge length to area ratio would then translate to a larger area on the substrate covered by the ferromagnetic edges relative to the diamagnetic bulk. In order to draw a comparison, we calculated the perimeter-to-area ratio $(\eta)$ of the flakes from each sample using the mean values obtained from each optical image. For this purpose, we obtained the perimeter by calculating the actual edge length, i.e., $l = l_o/\cos \phi$ and perimeter $P=3l$, using the mean values acquired previously. Here, $\phi$ corresponds to the vertex angle as illustrated in Figure~\ref{optical microscope images of flakes 700} (b).
Similarly, we find the area $A = l_o^2 \left(\sqrt{3} - 3\tan\phi\right)/4$, and the ratio of the perimeter-to-area $\eta =P/A$.

 \begin{figure} [ht!]
    \centering
    \includegraphics[width=0.75\linewidth]{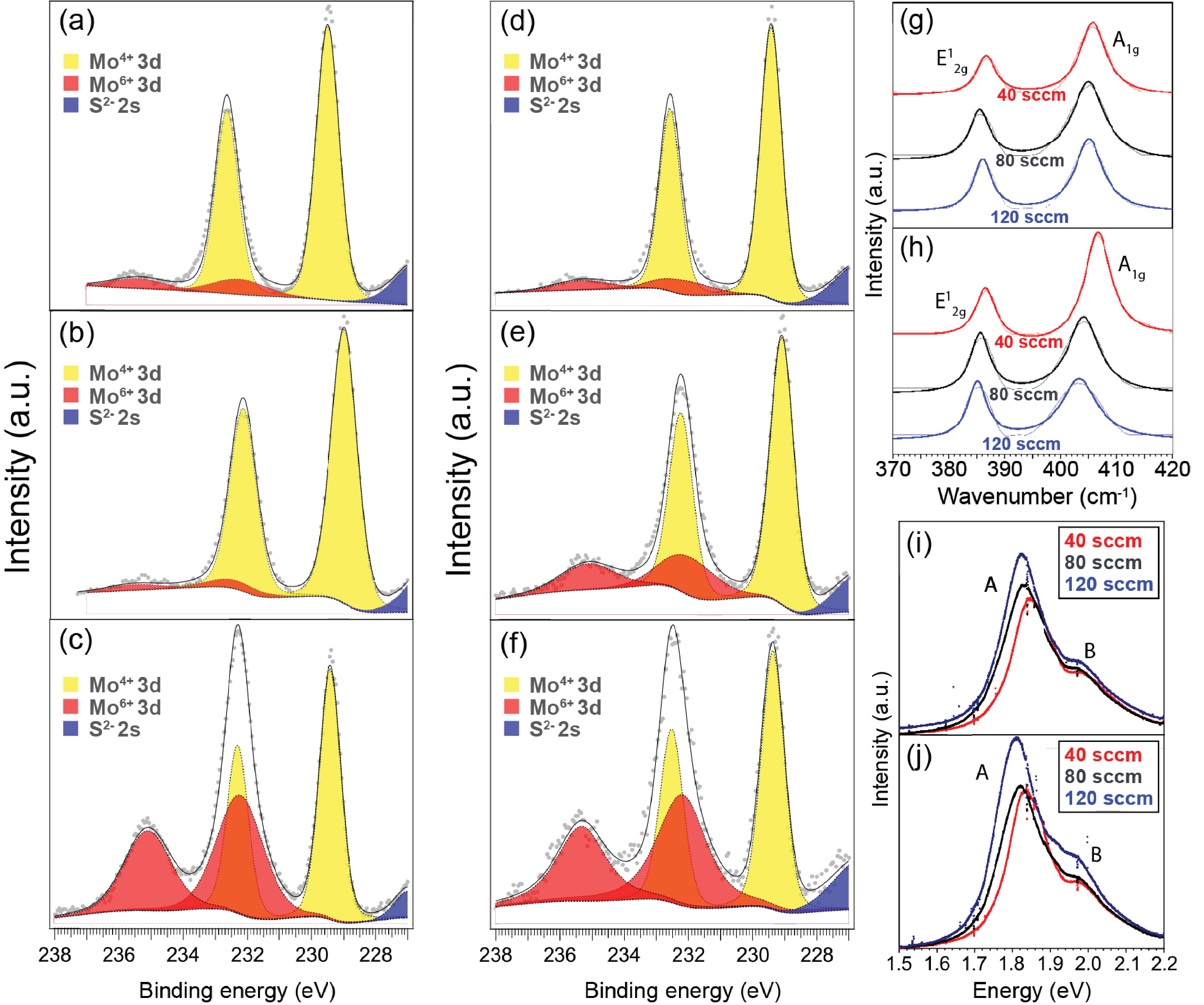}
    \caption{{XPS spectra of MoS$_{2}$ grown at 700$^{\circ}$C with Ar gas flow rate of (a) $120~$sccm, (b) $80~$sccm and (c) $40~$sccm from Mo 3d orbital. XPS spectra of MoS$_{2}$ grown at 800$^{\circ}$C with Ar gas flow rate of (d) $120~$sccm, (e) $80~$sccm and (f) $40~$sccm from Mo 3d orbital. Raman spectra of MoS$_{2}$ grown at (g) 700$^{\circ}$C and (h) 800$^{\circ}$C.
 Photoluminescence spectra of MoS$_{2}$ grown at (i) 700$^{\circ}$C and (j) 800$^{\circ}$C.}}
    \label{XPS PL Raman SEM}
\end{figure}

Figure~\ref{perimeter to area ratio 700 and 800} showcases the relationship for MoS$_{2}$ flakes grown at 700$^{\circ}$C and 800$^{\circ}$C. The dotted line represents ratios for flakes with the same edge length ($l_{o}$) with edges assumed to be straight for comparison. The ratio is influenced significantly by the flake sizes, while the ratio does not take into account the nucleation density; it was seen previously that the nucleation density followed a trend inversely related to the flake size. Hence, we note that the sample with the highest value for $\eta$ also has flakes growing closer to each other. Therefore, flakes with high value of $\eta$ growing close to each other must possess a high density of flake edges, given that other factors are considered as constants. Additionally, there is an appreciable increase in the ratio of ferromagnetic edges with respect to the diamagnetic bulk as the flakes curves inwards at high flow rates. The observed variations in flake morphology, particularly the high density of zigzag edges, provide a robust framework for enhancing ferromagnetic response. These structural features, induced by precise growth conditions facilitate spin polarization and magnetic coupling, making them crucial for studying defect-driven ferromagnetism in monolayer MoS$_2$, the main goal of this article.

\subsection{\label{sec:level2}Spectroscopic analysis}
To investigate the composition of our samples, we employed X-ray photoelectron spectroscopy (XPS). The findings from XPS provide a direct evidence of defect-state evolution with respect to varying growth conditions. The main features includes identifying the chemical states and binding energies of key elements in the samples, i.e, sulfur and molybdenum. The results are presented in Figure~\ref{XPS PL Raman SEM}, where each peak presents a doublet due to spin orbit splitting of the Mo 3d orbital. The yellow peaks corresponds to the the Mo\textsuperscript{4+} state bonded to S, while the red peak represents the Mo\textsuperscript{6+} state bonded to O. The blue peak from S\textsuperscript{2-} 2s only gives rise to a singlet due to absence of orbital angular momentum associated with s orbitals. We calculate the stoichiometric ratios of each sample using the relation:
\begin{equation}\chi= \frac{S}{Mo} = \frac{\frac{I_{S2p}}{\sigma_{S2p}}}{\frac{I_{Mo3d}}{\sigma_{Mo3d}}}, \label{eq:stoichiometric_ratio} \end{equation}

\noindent where $I_{S2p}$ and  $I_{Mo3d}$ are the integrated intensities of S 2p and Mo 3d peaks respectively. While $\sigma_{S2p}$ and $\sigma_{Mo3d}$ are the Hartree-Slater photoionization cross-sections at 1487$~$eV for S 2p and Mo 3d respectively \cite{scofield1976hartree}.

\begin{figure}[h!]
    \centering
    \includegraphics[width=1\linewidth]{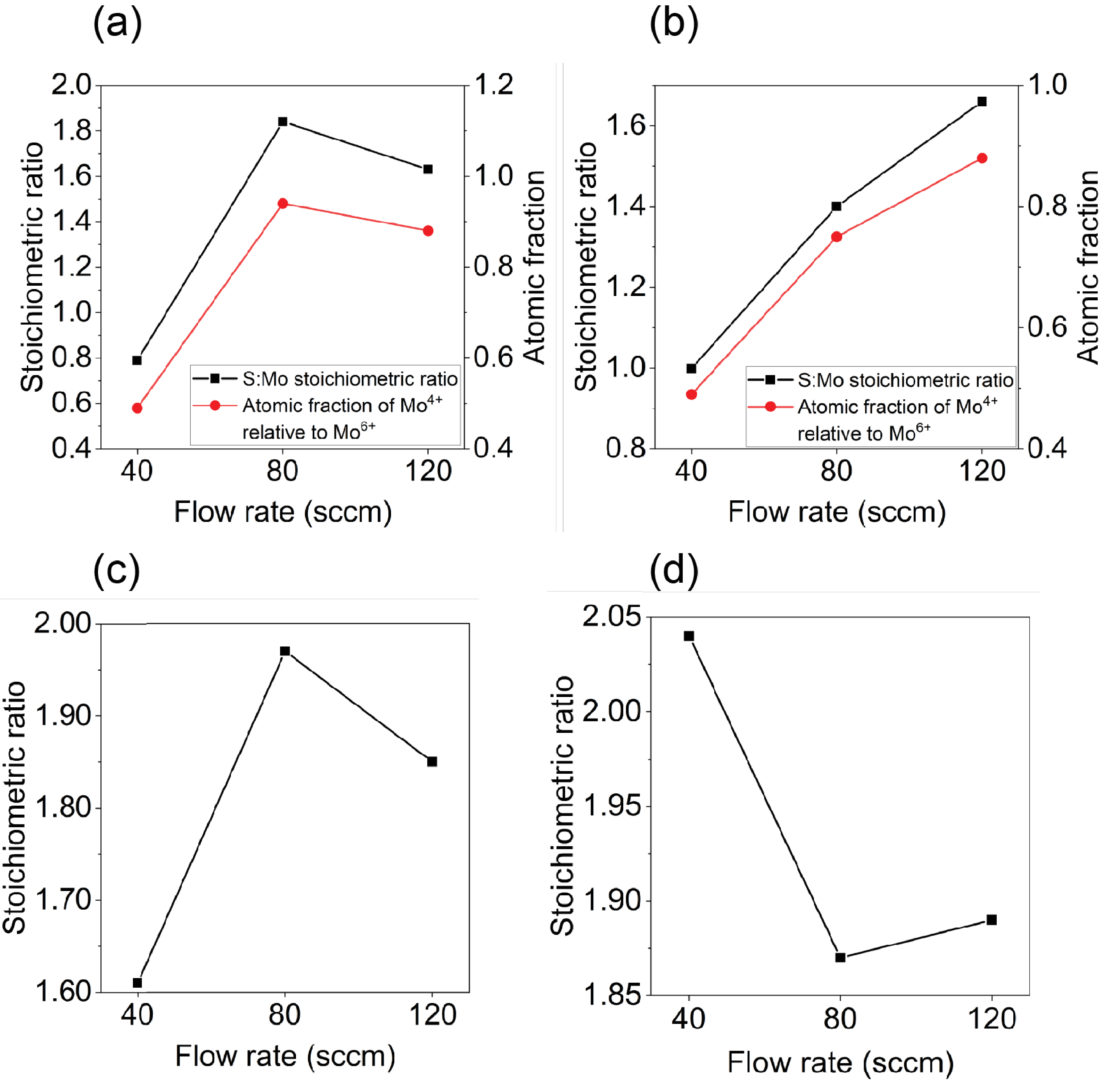}
    \caption{The black line represents the stoichiometric ratios $\chi$ of S:Mo in each sample synthesized at (a) 700$^{\circ}$C and (b) 800$^{\circ}$C. The red line corresponds to the atomic fraction of Mo\textsuperscript{4+} oxidation state with respect to Mo\textsuperscript{6+} oxidation state at (a) 700$^{\circ}$C and (b) 800$^{\circ}$C. Stoichiometric ratio of S:Mo\textsuperscript{4+} for each sample synthesized at (c) 700$^{\circ}$C and (d) 800$^{\circ}$C. }
    \label{XPS ratios}
\end{figure}

The ratio $\chi$ at each temperature increases as the amount of S supplied increases with increasing flow rate. Flakes grown at low flow rates were exposed to a high Mo environment which increased the probability of formation of S vacancies. These defects are susceptible to oxidation when exposed to air and can accelerate the aging process in MoS$_{2}$ flakes when exposed to moisture. Correspondingly, we note that the Mo\textsuperscript{6+} peak diminishes as the S supply at higher flow rate was increased, indicating that as vacancies were filled out, the rate at which oxidation occurs also decreased.
Deposition of unreacted MoO$_{3}$ particles could also contribute to the Mo\textsuperscript{6+} peak, however, it is less likely as we do not find such particles in the microscopic analysis further away from the Mo source where the XPS study was done. Moreover, we would expect particles deposited far away from the Mo source (and closer to the S source) to be at least partially sulfurized, and hence have a valency lower than 6+. Likewise, the Mo\textsuperscript{6+} peak is also reported in MoS$_{2}$ grown using MoO$_{2}$ precursor, leading us to believe that while contamination from unreacted particles may play a part, oxidation of S vacancies are the primary source of this peak \cite{somphonsane2023cvd}. The relative amount of Mo\textsuperscript{4+} state bonded to sulfur with respect to the Mo\textsuperscript{6+} state bonded to oxygen was used to estimate the amount of unoxidized sulfur vacancies. Consequently, we find that the vacancy sites in the flakes grown at 800$^{\circ}$C with a flow rate of $40~$sccm were completely passivated by oxygen; see Figure~\ref{XPS ratios} (d). Conversely, the samples grown at higher flow rates still consisted of sulfur vacancies which were not completely oxidized. Furthermore, the samples grown at 700$^{\circ}$C displayed a lower stoichiometric ratio between S:Mo atoms compared to their counterparts synthesized at high temperature despite the high S:Mo vapor concentration during growth. Due to a decrease in the mobility of atoms during growth at low temperatures, the sulfur atoms may have failed to fill in the vacancy sites completely, thus leaving behind a larger number of sulfur vacancies. At 80 sccm the Mo\textsuperscript{6+} peak diminishes significantly, and the S:Mo\textsuperscript{4+} stoichiometric ratio approaches the value 2. These results indicate that at low temperature, this optimal flow rate facilitates the reaction between the precursors to reach completion. Increasing the flow rate further causes the flake growth to be non-uniform with a higher amount of defects. Additional results from XPS are shown in the supplementary material.

CVD grown MoS$_{2}$ monolayers generally comprise of two types of sulfur vacancies V$_{S}$ and V$_{S2}$, with formation energies of 2.12$~$eV and 4.14$~$eV respectively \cite{hong2015exploring}. These vacancies form triangular shaped pits consisting of zigzag edges within the flake, with the pit triangle oriented in the direction opposite to that of the original flake. Since the V$_{S2}$ vacancies form larger pits, they are more prone to oxidation due to the presence of a larger number of dangling bonds in comparison to V$_{S}$ vacancies \cite{chen2018degradation}. We infer from this that the samples showing a higher degree of defect passivation primarily possess V$_{S2}$ vacancies. This accounts for the lower amount of unfilled vacancies present in flakes grown at high temperatures where the growth conditions are sulfur deficient. In contrast, at lower temperature the vacancies are expected to be predominantly of V$_{S}$ type, therefore they are not completely oxidized and instead are filled up by sulfur atoms as the flow rate increases.

\begin{figure}
    \centering
    \includegraphics[width=0.83\linewidth]{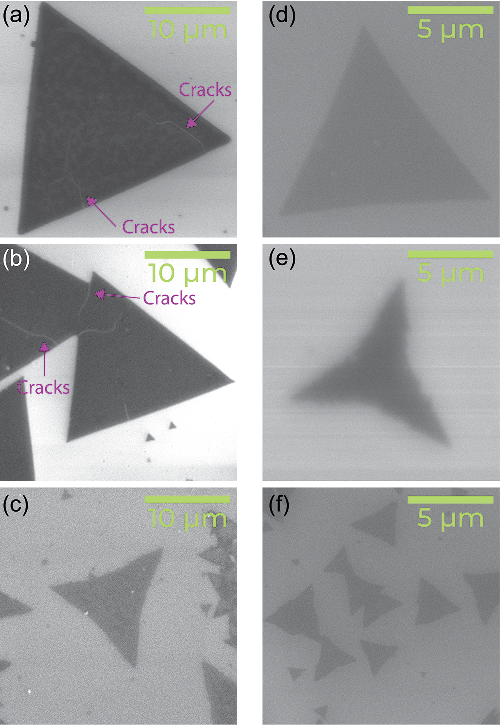}
    \caption{Scanning electron microscope images of MoS$_{2}$ grown at 800$^{\circ}$C with a flow rate of (a) $40~$sccm, (b) $80~$sccm and (c) $120~$sccm. Scanning electron microscope images of MoS$_{2}$ grown at 700$^{\circ}$C with a flow rate of (d) $40~$sccm, (e) $80~$sccm and (f) $120~$sccm.}
    \label{SEM}
\end{figure}

Scanning electron microscope (SEM) images revealed that flakes grown at high temperatures appear to be cracked, particularly at low flow rates, refer to Figure~\ref{SEM}. It is known that the chalcogen vacancies act as sites for stress concentration thereby initiating cracking due to thermal contraction of the flakes upon rapid cooling \cite{hao2017orientation}. These cracks give rise to new anti-symmetric energy favorable zigzag edges, which are S-terminated on one side and Mo-terminated on the other side \cite{jung2018interlocking,wang2015fracture}. These cracks can substantially enhance the density of zigzag edges to a great extent and contribute to the overall magnetization of the sample. On the contrary, the cracks were largely absent in the flakes grown at 700$^{\circ}$C leading us to believe that rapid cooling from a high temperature was the primary cause of their formation.

Additionally, Raman spectroscopy was used to identify the number of layers of the MoS$_{2}$ flakes grown in each sample. The Raman spectrum of MoS$_{2}$ exhibits two characteristic modes for identification, namely the in-plane vibrational mode $E^{1}_{2g}$ and the out of plane vibrational mode $A_{1g}$. The frequencies of these peaks can help provide insights into the number of layers, strain and defects in atomic layer MoS$_{2}$. We make use of the difference in the peak frequencies to determine the number of layers in our samples. This difference is observed to be $\Delta k~\leq~20~cm^{-1}$ for all samples, suggesting that they are monolayer \cite{mccreary2018and}, see Figure~\ref{XPS PL Raman SEM}. Notably, when the flow rate is increased, the $A_{1g}$ peak broadens and undergoes a blue-shift, with its intensity diminishing at each step. On the contrary, we noted little change in the $E^{1}_{2g}$ across different samples suggesting that the strain effects are negligible. The broadening of linewidth along with diminishing intensity of $A_{1g}$
Raman mode, relative to the $E^{1}_{2g}$, as flow rate increases, provides evidence for increased electron doping at high flow rates. The $A_{1g}$ phonon is understood to exhibit a strong electron-phonon coupling which is influenced greatly by electron doping \cite{chakraborty2012symmetry}.

\begin{figure*}[ht!]
    \centering
    \includegraphics[width=0.8\linewidth]{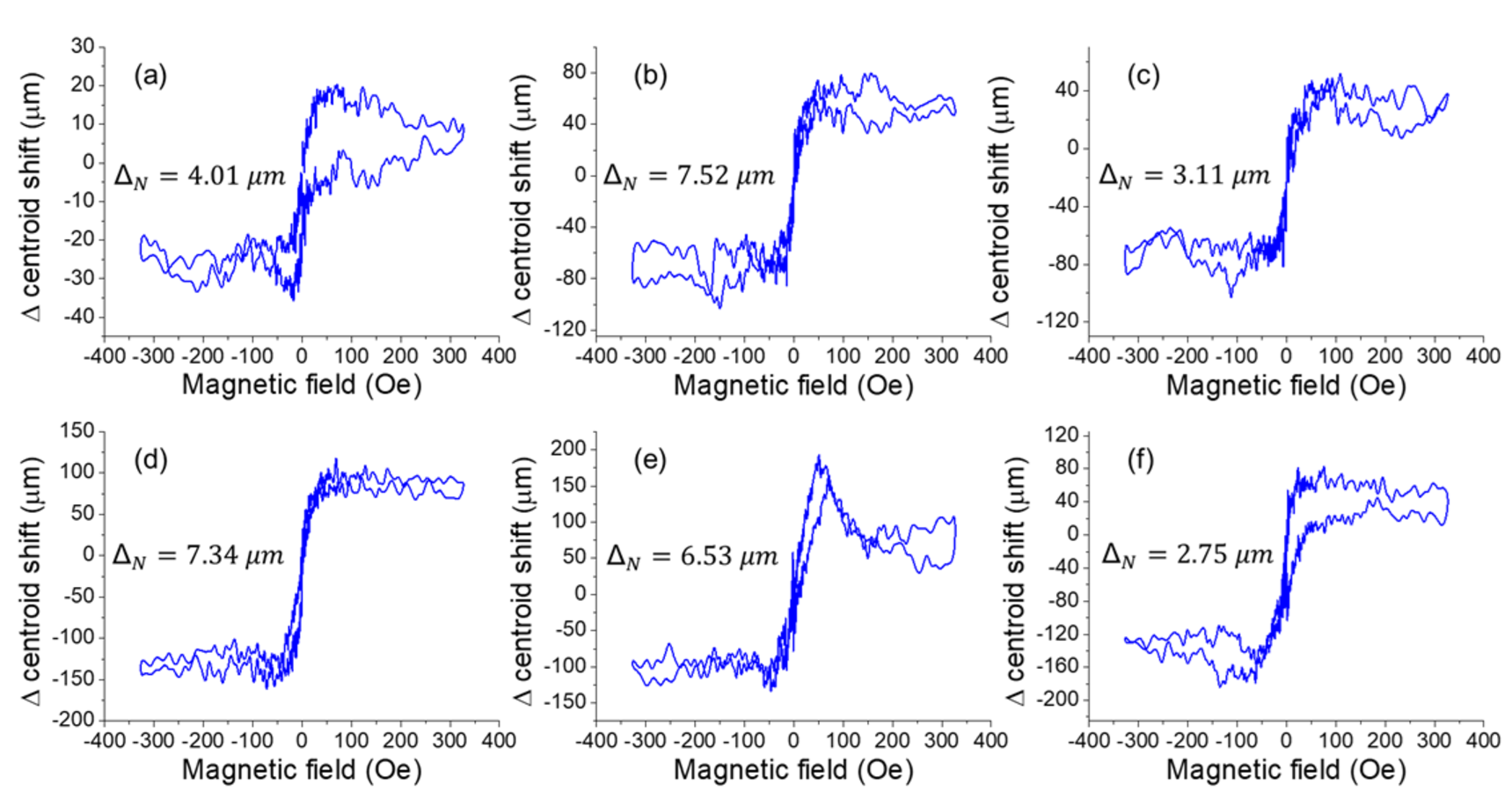}
    \caption{SHEL-MOKE showing centroid shift ($\Delta$) as a function of applied external magnetic field in longitudinal configuration for samples grown using CVD. For flakes synthesized at 700$^{\circ}$C with Ar gas flow rates of (a) $40~$sccm, (b) $80~$sccm and (c) $120~$sccm. Furthermore, for MoS$_2$ flakes grown at 800$^{\circ}$C with Ar gas flow rates of (d) $40~$sccm, (e) $80~$sccm and (f) $120~$sccm. In each graph $\Delta_N$ is the normalized centroid shift for the respective sample.}
    \label{SHEL MOKE centroid hystersis}
\end{figure*}
Finally the photoluminescence (PL) spectra, shown in Figure \ref{XPS PL Raman SEM}, each sample present two peaks arising from direct band transitions between $\sim$1.80$~$eV\textendash1.85$~$eV, which varies as a function of the gas flow rate during growth, and $\sim$1.97$~$eV, corresponding to the A and B exciton peak respectively. The two peaks arise due to spin splitting in the valence band, originating from spin orbit coupling. It is observed that the samples grown at progressively higher flow rates exhibit increased PL intensities, with the peak shifting to lower energies. This behavior is attributed to the electron doping occurring due to excessive S physisorption on the surface of MoS$_{2}$ at higher flow rates. Electron doping introduces charge carriers, enhancing exchange interactions between localized magnetic moments at edges and defect sites in MoS$_{2}$ \cite{bao2022brief}. Moisture can penetrate between the flakes and substrate resulting in quenched PL intensity in highly oxidized flakes. Moreover, electron doping increases the contribution of trions to the A peak causing it to be red shifted \cite{zhu2023room}. Likewise, excess S lowers the likelihood of formation of point defects and ensures greater crystal quality diminishing non radiative recombination through defects, thus resulting in a higher PL intensity. In short, PL spectroscopy confirms the impact of defect engineering on the optical and magnetic properties of MoS$_{2}$, further supporting the insights gained from our Raman and XPS analysis.

\subsection{\label{sec:level2}SHEL-MOKE}
The detailed spectroscopic analysis provides strong evidence that defect states from CVD grown flakes are associated with creating magnetic moments. Raman analysis provides evidence for increased electron doping at high flow rates facilitating electron-phonon coupling. The electron doping is further supported by PL analysis, where the trions contribute towards red-shift of peaks that is directly linked to changes in exchange interactions at defect sites. These interactions amplify spin polarization and consequently the magnetic response. Furthermore, XPS results confirm the presence of sulfur vacancies and Mo$^{+6}$ which are well known to be associated to magnetic moments due to unpaired electrons. Together, these spectroscopic analysis provide a comprehensive understanding of defect states, consistent with literature \cite{park2024strong,cai2015vacancy,tongay2012magnetic,zhou2018robust}, showing that these are polarization-active sites that are indeed magnetic in nature. Such samples are ideal candidates to observe spin Hall effect of light (SHEL) as a function of external magnetic field.

The SHEL has been a well known tool in optics to measure subtle polarization and magnetic changes in different materials, mainly because the observed output of SHEL is dependent on the polarization of incident light, even minute change in polarization upon reflection from the sample can drastically affect the output signal. Panda \emph{et al.} have used SHEL based magneto-optic Kerr effect technique to investigate the magnetic and optical properties of ultra-thin films of permalloy and permalloy-MoS$_{2}$ heterostructures \cite{panda2022high}. Here, we use SHEL-MOKE centroid shift measurements to provide qualitative insights into the ferromagnetic properties of thin films. A controlled experiment to verify the SHEL-MOKE response is performed on a standard magnetic sample, permalloy (NiFe), and compared to it's MOKE response (refer to supplementary material). Subsequently, we measure SHEL-MOKE for the CVD grown samples. The synthesis conditions, particularly temperature and gas flow rate, significantly influence the magnetic domain defect density, thereby modulating the ferromagnetic response.

In Figure~\ref{SHEL MOKE centroid hystersis} the centroid shift ($\Delta$) is plotted as a function of applied magnetic field for all CVD grown samples. The shift is a measure of the SHEL splitting, which is sensitive to the ability of a sample to affect the polarization of laser upon reflection from the sample surface. All samples show a hysteresis like response when placed in longitudinal magnetic field, indicating a ferromagnetic signature. For comparison we calculate the normalized centroid shift as
\begin{equation}
\Delta_N=\bigg|\frac{\Delta_{max}-\Delta_{min}}{\Delta_{max}}\bigg|.
\end{equation}

\noindent The normalized centroid shift, $\Delta_N$, is used as a measure to compare the magnetic response of all samples. For samples synthesized at 800$^{\circ}$C, the overall centroid shifts are larger compared to samples at 700$^{\circ}$C. The 800$^{\circ}$C samples particularly at $40~$sccm and $80~$sccm displayed noticeable cracks, that further contribute as defect edge states. Moreover, the presence of pits arising from V$_{S2}$ vacancies in samples synthesized under S deficient condition could give rise to even more zigzag edges. The close proximity of the edges arising from these cracks and pits are likely to facilitate strong ferromagnetic coupling with the adjacent edges emerging from defects. In addition, the presence of Mo\textsuperscript{6+} state could also be a source of enhanced ferromagnetism \cite{singh2023highly}, occurring from exchange interaction between different oxidation states of Mo. Wei \emph{et al.} have also shown through DFT calculations that oxidized MoS$_2$ exhibits a net magnetic moment, primarily due to the density of states of oxygen near the Fermi level \cite{xie2022plasma}. For $120~$sccm, $\Delta_N$ decreases, which can be associated to the fact that at high flow rate the edge type changes to S-terminated, as evident from SEM and edge angle measurements (see Figure~\ref{edge length and flake angle}). This S-terminated edge type is linked with much lower magnetic response compared to Mo-terminated edges \cite{wang2017edge,li2008mos2}. This outcome is reasonably aligned with our analysis of defect densities and enhanced magnetic interactions discussed in spectroscopy results.

The flakes synthesized at 700$^{\circ}$C have a notably smaller centroid shift indicating weak magnetization, mainly because all samples grown at 700$^{\circ}$C have S-terminated edges. The presence of S-terminated is validated through SEM images, Figure~\ref{SEM}, and our measurements of edge angles, Figure~\ref{edge length and flake angle}. The small shift in $\Delta_N$ at $40~$sccm is expected due to the existence of S vacancies which have a detrimental effect on the magnetic moment originating from S-terminated edges \cite{wang2017edge}. For sample grown at $80~$sccm, $\Delta_N$ shows a comparative increase, whereas it $\Delta_N$ decreases for sample grown at $120~$sccm as expected. The anamolous magnetic behavior of sample at $80~$sccm flow rate can be associated with higher flake density and a much lower edge angle compared to the edge angle of sample grown at $120~$sccm.

The hysteresis patterns observed in our SHEL-MOKE measurements corroborate the predicted responses from our spectroscopic analysis, which highlighted the role of defects in inducing and diminishing spin-like behavior. This response cannot be elicit from a conventional MOKE experiment due to the ultrasmall signal-to-noise ratio and needs amplification through the post-selection criterion embedded in the quantum weak value scheme. The supplementary material provides null results from control experiments, which forms an important benchmark validating the efficiency of the weak measurement technique. This trend can be critical for applications that rely on defect-induced magnetism in two-dimensional materials. As a result, the magnetic response of CVD grown thin layer MoS$_2$ is successfully probed using the highly sensitive SHEL-MOKE experiment. We conclude that even though Mo edge states are the primary cause of magnetic moments in these samples, parameters like flake density, presence of crack and pits and post-oxidation of vacancies can all affect the collective magnetic signal. Through our theoretical and experimental analysis, we validate that SHEL-MOKE not only reveals magnetic response from defect-engineered MoS$_2$ flakes but also proves its capability as a powerful tool for detecting polarization dependent changes in unique systems.

\section{\label{sec:level1}Summary}
In this work, we present ferromagnetic behavior in monolayer CVD-grown MoS$_{2}$. The morphologies and edge terminations of the samples produced under various growth conditions have a distinct effect on the magnetic characteristics. We present the efficiency of quantum weak measurements by using SHEL-MOKE to experimentally probe ultra-small magnetic response from two-dimensional system which are unobservable with conventional MOKE. It is observed that Mo-terminated edges produce a greater magnetic moment than S-terminated edges, and our results are consistent with earlier research that connected ferromagnetism to zigzag edge geometries. Furthermore, we found that whereas extending edge lengths through engineering flake curvature can enhance the S-terminated edge state density, it eventually decreases the total magnetization. Crucially, we discover that the flakes' internal fissures significantly improve MoS$_{2}$'s magnetic characteristics by forming additional zigzag edges that enhance the magnetic behavior. We also show SHEL-MOKE to be a useful avenue to explore small but useful information about the quantum nature of novel materials, particularly atomic crystals of semiconductors.

\section*{Supplementary Material}
The supplementary material includes experimental setups for MOKE and SHEL-MOKE. Additional null experiments from standard samples have been shown to verify the reliability of measurement schemes. X-ray photoelectron spectra for carbon 1s and sulfur 2p are given for reference.

\section*{ACKNOWLEDGMENTS}
Wardah acknowledges receipt of Syed Babar Ali Research Award (SBARA) funding. Authors would like to thank Muhammad Rizwan, Khadim Mehmood and Muhammad Ayyaz for technical support, Shahzad Akhtar Ali for help in Raman spectroscopy, and Shirin Abbas for help in the initial stages of the synthesis of atomic flakes.
\section*{Data Availability Statement}
The data that support the findings of this study are available within the article [and its supplementary material].

\end{document}